\documentclass{optica-article}

\journal{opticajournal}
\articletype{Research Article}

\usepackage{amsmath,amsfonts}

\begin{document}

\title{DO-CGI: deep-optimized illumination patterns for computational ghost imaging at low sampling ratios}

\author{Mor Hale,\authormark{1} Ofir Lindenbaum,\authormark{1} and Eliahu Cohen\authormark{1,*}}

\address{\authormark{1}Faculty of Engineering, Bar-Ilan University, Ramat Gan 5290002, Israel}

\email{\authormark{*}eliahu.cohen@biu.ac.il}

\begin{abstract*}
Computational ghost imaging (CGI) reconstructs objects from known illumination patterns and bucket-detector measurements, but quality deteriorates at low sampling ratios (SRs). We present a deep-learning framework that optimizes grayscale diffuser patterns before reconstruction. In simulations using CIFAR-10 and MNIST images with Split Bregman reconstruction, the learned patterns outperform random patterns in peak signal-to-noise ratio and structural similarity, including at SRs below 5\%. Patterns trained on CIFAR-10 also transfer to MNIST and remain effective under moderate perturbations of the sensing matrix. These results support learned pattern design as a route to fewer CGI measurements.
\end{abstract*}

\section{Introduction}
Computational ghost imaging (CGI) is an advanced imaging technique that captures information about an object by illuminating it with a series of structured light patterns (diffusers) and subsequently reconstructing the image through sophisticated restoration algorithms. Utilizing a single-pixel detector to capture photons that have interacted with the object, CGI offers substantial advantages over traditional imaging methods. These advantages include improved resolution, enhanced sensitivity, reduced measurement time, lower illumination power, broader spectral range, and cost efficiency \cite{edgar2019principles,gibson2020review12}.

More specifically, CGI holds significant potential for X-ray imaging \cite{sefi202420}, which is used in fields ranging from basic science and industry to medicine and nondestructive testing. X-ray instruments face practical tradeoffs among resolution, contrast, acquisition time, and radiation exposure, together with limited availability of high-quality X-ray optics. CGI combined with compressed sensing has therefore been investigated as a route to dose-efficient X-ray imaging \cite{huang2024deep}. In the present simulation study, however, a lower number of measurements should not be identified automatically with a lower dose: that conclusion also requires control of the exposure per pattern and the total photon budget.

An important challenge in CGI is to shorten the acquisition time without unduly compromising image quality. When the sampling ratio (SR) is below 100\%, the linear measurement system is underdetermined and reconstruction depends on prior structure, regularization, and the properties of the illumination patterns.
Numerous algorithms designed to tackle ill-posed optimization problems can be employed for image reconstruction in CGI. Examples include the Split Bregman algorithm  \cite{yang2020modifiedsplitBregman}, orthogonal matching pursuit (OMP) algorithm in \cite{tropp2007OMP}, alternating direction method of multipliers (ADMM) algorithm in \cite{boyd2011ADMM}, conventional Split-Bregman (CSB) algorithm in \cite{goldstein2009CSB}, projected Landweber regularization with guide filter (PLRG) algorithm in \cite{huang2018PLRG} and Total Variation Augmented Lagrangian Alternating Direction (TVAL3) algorithm in \cite{li2013TVAL3}.

A broadly used set of techniques to address a diminished SR is proposed by the theory of compressed sensing (CS) \cite{katkovnik2012compressive}. Similarly to previously mentioned methods, CS operates under the assumption that the target image exhibits sparsity in a particular basis.The goal is to recover this sparse signal using a limited number of measurements. Although these methods have been studied before and have shown some success, their effectiveness is often limited, especially when dealing with low SR  \cite{katz2009compressive, zhang2021computational, zhang2022efficient}.

In recent years, learning-based methodologies have gained considerable traction as effective solutions to a diverse range of challenges spanning various domains, including optical imaging. The increased interest in employing deep learning (DL) for computational imaging tasks has been particularly noteworthy \cite{rizvi2019Fourier, shimobaba2018computationaldeep,wu2020denoising,zhu2020Ynet,wang2019learning,wang2022GIDC}. Similarly, DL tools are employed within CGI via two primary routes: denoising of low-quality images which are reconstructed from conventional algorithms \cite{shimobaba2018computationaldeep,lyu2017deepghost, he2018ghostdeep,bian2020residual,rizvi2020deepghost,wang2022farfieldGIDC,li2023singlepixelUCAN} and direct image reconstruction from raw data collected by the detector \cite{wang2019learning,wu2020denoisingDDANet,liu2021computationalUNNCGI,zhu2020ghostYnet}. 

The first scenario often arises when dealing with noisy or distorted images generated through traditional processes. Deep learning techniques come into play to clean up these images, enhancing their quality and making them more suitable for various applications such as medical imaging, surveillance, or any field where image fidelity is paramount. Among these approaches within computational imaging, stands out the deep convolutional autoencoder network (DCAN) \cite{rizvi2019Fourier}. During training, the DCAN takes noisy input images and compares its reconstructed outputs against their corresponding clean ground-truth counterparts, thereby quantifying the discrepancy between them. The model is trained to attenuate noise and artifacts while preserving semantically meaningful image details. However, the employment of DCAN often relies heavily on pre-training strategies with existing data, which can be time-consuming and are not always suitable to CGI due to the lack of sufficient experimental data pertaining to a variety of objects at various conditions. Another prominent DL-based method for this challenge is ghost imaging with a deep neural network constraint (GIDC), introduced in \cite{wang2022farfieldGIDC}. Notably, this network does not require a large dataset for training; it can be trained using just one image. However, there are some drawbacks to consider. GIDC requires an initial low-quality reconstruction, and the network must undergo full retraining for each object, adding a layer of complexity to the process.

As discussed above, except for using deep learning to enhance low-quality reconstruction, there are additional attempts to improve CGI quality using direct reconstruction of images from bucket signals using supervised learning of large datasets via neural networks. These learning-based methods represent a significant advancement in image reconstruction, where deep neural networks are utilized to directly convert raw signal data into meaningful and visually coherent images. This method bypasses the intermediate steps of traditional algorithms, offering a more efficient process for image reconstruction. 

In most of the DL-based reconstruction methods discussed above, a fixed random diffuser set is used during training and must be reproduced during testing. Pattern design therefore remains relevant even when reconstruction is learned. Established alternatives include Hadamard, Fourier, discrete cosine transform (DCT), and wavelet patterns, which can outperform random patterns in suitable settings \cite{zhao2023comparison}.


Here we take a complementary route: deep learning is used to optimize the diffuser set before reconstruction, while image recovery is performed by a standard compressed-sensing algorithm. The objective is to learn patterns that improve reconstruction quality at low SR and to test whether patterns trained on one image domain retain utility in another. Throughout this article, ``optimized'' means optimized with respect to the stated network, loss function, training set, and training procedure; it does not imply a global optimum. The quantitative baseline in the present study is a random diffuser ensemble. Comparisons with Hadamard, Fourier, DCT, wavelet, or other optimized pattern families remain necessary before claiming superiority over established structured designs.

\section{Background}

 The ghost imaging (GI) system comprises two distributed light beams: the test and reference beams. In the test beam, light illuminates the object and is collected into a bucket measurement. The reference beam, on the other hand, travels freely to a high-resolution detector without interacting with the object. Image reconstruction is accomplished through correlation measurements between the signals of the two light beams.

\begin{figure}[htbp]
    \centering
        \includegraphics[width=0.75\linewidth]{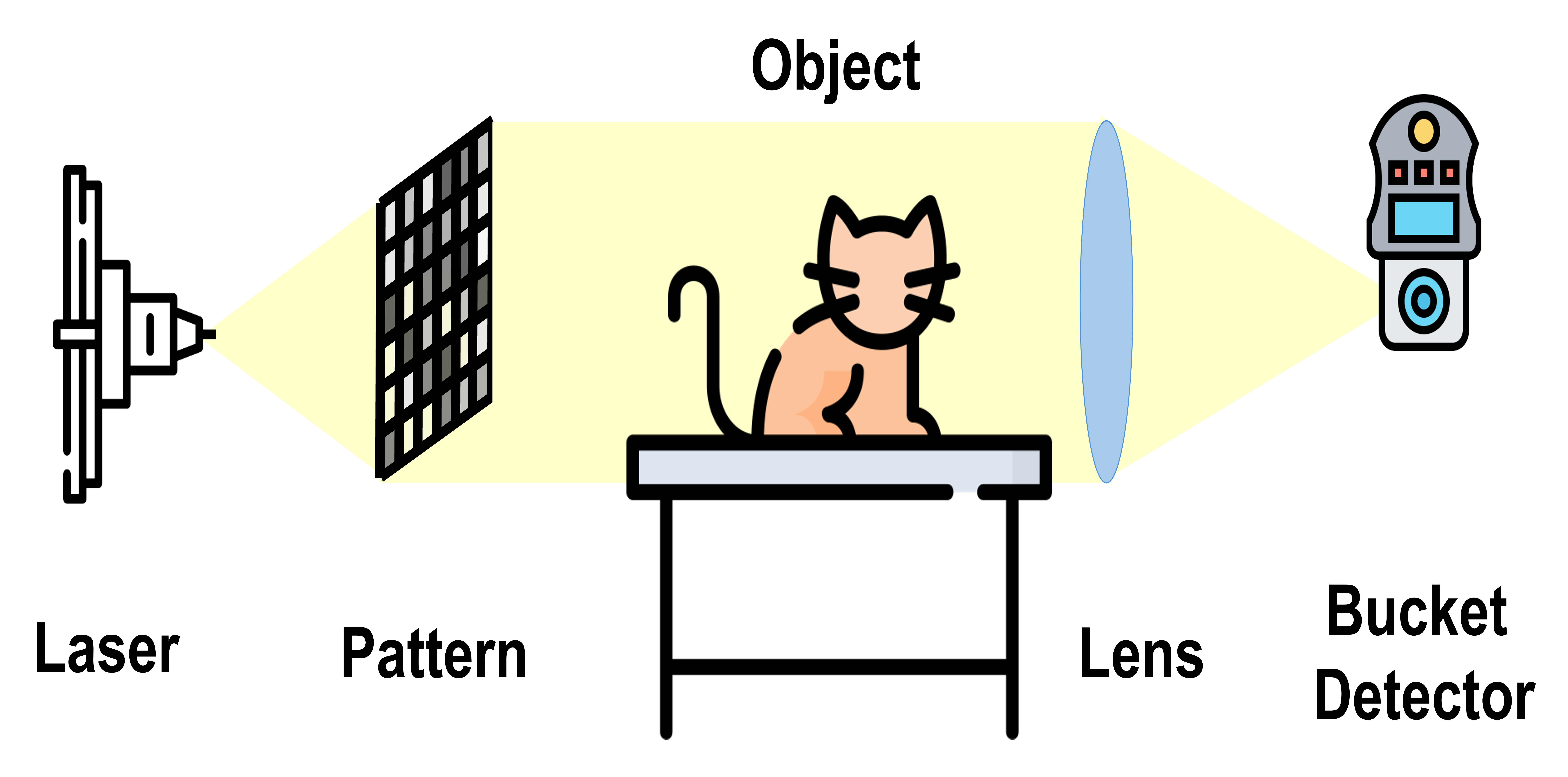}
    \caption{CGI setup. The laser source emits monochromatic light, which undergoes modulations using a computer-controlled SLM-generated diffuser. A lens positioned after the object precisely focuses the light onto a bucket detector, measuring the total intensity of the illuminated light.}
    \label{fig: CGI Setup}
\end{figure}

The reference beam is no longer required in computational ghost imaging (CGI). Instead, the image is obtained by calculating the correlation between the test beam intensity and the known diffuser displayed on the spatial light modulator (SLM). This CGI setup is illustrated in Fig. \ref{fig: CGI Setup}.

Without prior information, unique recovery of a generic \(N\)-dimensional object typically requires at least \(N\) independent linear measurements. Compressed-sensing CGI instead exploits object structure, such as sparsity or low total variation, and can therefore produce useful reconstructions with \(M<N\). We define the sampling ratio as \(\beta=M/N\); values below 100\% correspond to deliberately underdetermined acquisition, whose success depends on the object class, pattern ensemble, noise, and reconstruction prior.

We describe the measured bucket-signal vector using
\begin{equation}
\label{eq:measurement_model}
\boldsymbol{y}
=
(\boldsymbol{A}+\Delta\boldsymbol{A})\boldsymbol{x}
+\boldsymbol{n},
\end{equation}
where \(\boldsymbol{A}\in\mathbb{R}^{M\times N}\) is the nominal pattern matrix, \(\Delta\boldsymbol{A}\) represents pattern-generation or calibration error, and \(\boldsymbol{n}\) represents additive bucket-detector noise. Given the calibrated matrix, the reconstruction problem is written as

\begin{equation}
\label{eq:optimization_problem}
\min_{\boldsymbol{x} \in \mathbb{R}^{N}}
\|\boldsymbol{A}\boldsymbol{x}-\boldsymbol{y}\|_2^2
+\lambda R(\boldsymbol{x}).
\end{equation}

Here, \(\boldsymbol{y}\in\mathbb{R}^{M}\) is the measured signal and \(\boldsymbol{x}\in\mathbb{R}^{N}\) is the object transmission profile, with \(N\) pixels. The regularizer \(R(\boldsymbol{x})\) can take different forms; here we use total variation, which favors piecewise-smooth reconstructions by penalizing large intensity gradients.

To design the diffuser, we employed deep neural networks (DNNs), which we trained to optimize the illumination pattern, with the goal of improving the image reconstruction quality at low SRs.

In recent years, deep neural networks (DNNs) have revolutionized the field of imaging, providing powerful tools for tasks such as image reconstruction, denoising, and super-resolution. Among these, representation learning has emerged as a pivotal approach to automatically discover meaningful features from raw data, reducing the need for manual feature engineering. Autoencoders (AEs), a well-established method in representation learning, transform data into a compact latent space through an encoder and then reconstruct it via a decoder. This approach has been extensively used in tasks such as dimensionality reduction, anomaly detection, and image denoising.

An alternative to traditional representation learning methods is implicit neural representations (INRs), which encode complex structures, such as images or 3D shapes, using continuous functions learned by neural networks. Typically implemented with multi-layer perceptrons (MLPs), INRs map spatial coordinates to scalar values, offering a continuous, highly scalable representation that requires minimal storage. INRs are particularly effective for applications like shape reconstruction and novel view synthesis, where high-quality outputs are essential.

In this work, we employ a neural network to map a noise vector to a matrix, which is combined with an image dataset to simulate computational ghost imaging (CGI). The resulting vector representation is subsequently used in a reconstruction process based on the split Bregman algorithm. The loss function for training is defined as the mean squared error (MSE) between the reconstructed and original images:

\begin{equation}
\textit{MSE} = \frac{1}{n} \sum_{i=1}^{n} (I_i - \hat{I}_i)^2,
\end{equation}

where \(I_i\) is the original image and \(\hat{I}_i\) is the reconstructed image.

Although our approach does not strictly align with traditional AE or INR paradigms, it draws on elements of both. The neural network functions as an encoder by capturing data representations in its weights, while the reconstruction process acts as a form of decoding, refining the network's output. This integration of neural networks and reconstruction algorithms enables an efficient encoding-decoding pipeline tailored to optimize image reconstruction quality under constrained conditions.

\section{Methodology and Implementation Details}

\begin{figure}[htbp]
\centering
    \includegraphics[width=\linewidth]{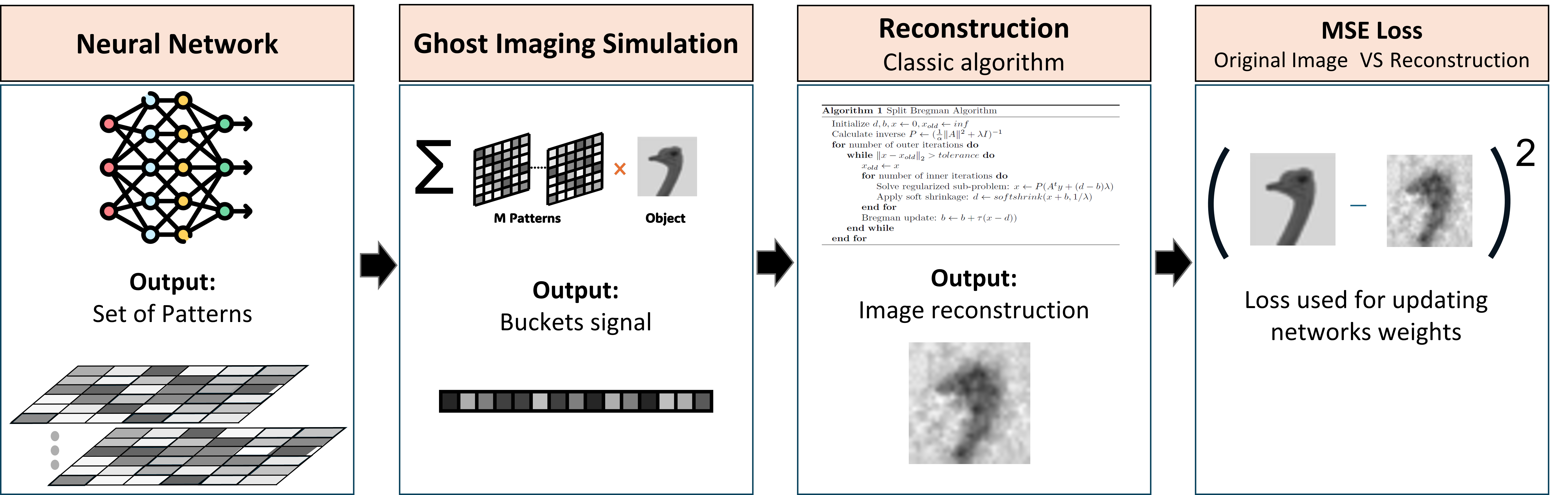}
    \caption{Overview of the training process. The process begins with the neural network (leftmost panel), which receives a noise vector as input to generate optimized imaging patterns. These patterns are then applied through our code simulating the GI process, as shown in Fig. \ref{fig: CGI Setup} (second panel), to produce bucket signals. In the reconstruction phase (third panel), the Split Bregman algorithm is used to reconstruct the image from the bucket signals. Finally, mean squared error (MSE) loss (rightmost panel) is calculated by comparing the reconstructed image to the original, and this loss guides the adjustment of the network's weights.}
    \label{fig: TrainingProcess}
\end{figure}

In this study, we develop a neural network that generates a set of diffuser patterns optimized for image reconstruction using Computational Ghost Imaging (CGI). The proposed framework is illustrated in Fig. \ref{fig: TrainingProcess}. The generated patterns are utilized to simulate bucket signals, with CIFAR-10 images serving as the input objects. These bucket signals, combined with the generated diffuser patterns, are input into a reconstruction algorithm to perform image reconstruction. The quality of the reconstructed images is assessed using the Mean Squared Error (MSE) between the reconstructed and original images, which also functions as the loss criterion during the network training process.

The neural network used to generate the optimized diffusers has a fully connected architecture with four layers. The first three layers contain 128, 256, and 512 neurons, respectively, and the final layer generates \(M\) patterns with the same pixel dimensions as the object. Each hidden layer incorporates batch normalization and a ReLU activation. A sigmoid output constrains every pattern value to \([0,1]\), so the reported masks are nonnegative grayscale transmission patterns. The simulations do not test binary quantization, spatial-light-modulator calibration, or an experimentally matched photon budget; those implementation constraints require separate validation.

\begin{figure}[htbp]
\centering
        \includegraphics[width=\linewidth]{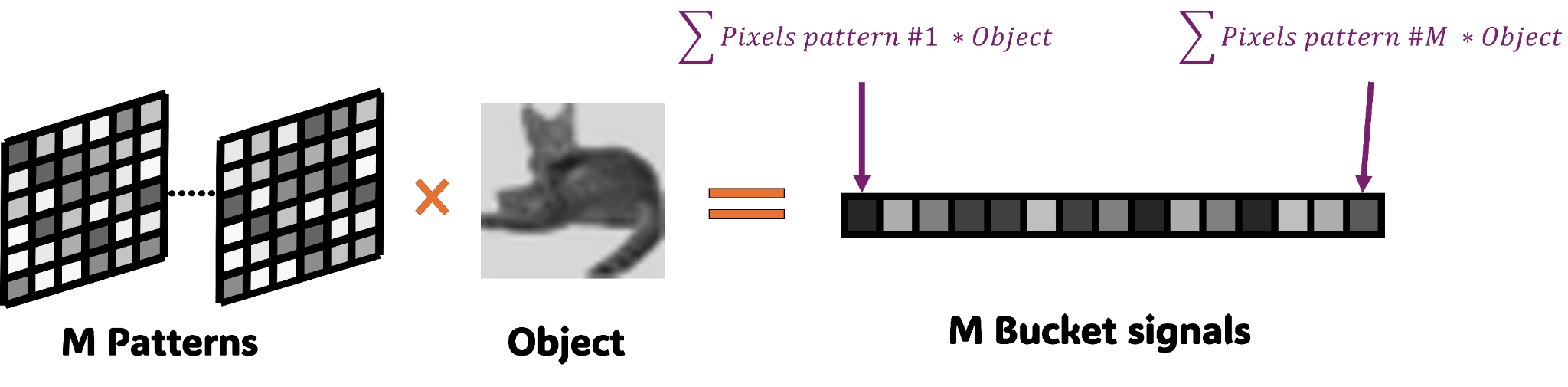}
    \caption{CGI simulation. The original image is multiplied by $M$ masks to generate $M$ transmitted light fields. Each transmitted light field is then summed to create a single bucket measurement. The final outcome is an $M$-length vector representing the bucket signal vector.}
    \label{fig: GI simulation}
\end{figure}

In our simulation, we utilized the approach presented by \cite{wang2019learning}, which is specifically designed for computational ghost imaging (CGI), to generate the bucket signals. This method, illustrated in Fig. \ref{fig: GI simulation}, allowed us to simulate the CGI process without the need for physical experiments.

The bucket signals were processed using a standard compressed-sensing reconstruction technique, the Split Bregman method with total-variation (TV) regularization. This iterative method addresses Eq. \ref{eq:optimization_problem} through alternating regularized updates, shrinkage, and a Bregman update. The parameters \(\alpha\) and \(\lambda\) control the data and penalty terms, while \(\tau\) scales the Bregman update. We set \(\alpha=\lambda=\tau=1\), used a tolerance of \(10^{-10}\), and performed one inner and one outer iteration. All quantitative results in this article use this reconstruction procedure. Although the pattern-learning architecture could in principle be paired with another differentiable or unrolled reconstruction method, performance with other algorithms is not demonstrated here.

After each batch reconstruction, the neural network updates its weights by minimizing the MSE between the original and reconstructed images, refining the diffuser progressively over each iteration.

For training, we used a randomly selected subset of 80 images from the CIFAR-10 dataset, with an additional 600 images reserved for testing. This choice reflects the practical constraints of computational imaging experiments, where obtaining a large variety of images of real objects if often challenging and time-consuming. The use of a small dataset is therefore necessary to ensure that the training process remains feasible when working with actual object images collected in laboratory experiments. Despite the small size of the training set, our approach enables the DNN to perform significant learning leading to good restoration results, hence demonstrating its effectiveness even under constrained conditions. Additionally, this approach accommodates limited computational resources and supports efficient retraining across multiple cycles.

The framework is adaptable for generating diffusers at various sampling ratios (SR), though retraining is required for each SR. Consistent hyperparameters were used across all SRs: the training procedure consisted of 10 epochs with a learning rate of 0.1, followed by 30 epochs at 0.01 and the final 5 epochs at 0.001. The batch size was set to 2, with each image sized at $32\times 32$ pixels, and a weight decay of \( 5 \times 10^{-7} \) was applied.

\noindent\textbf{Split Bregman update.}
The implemented update can be summarized without an additional algorithm package:
\begin{enumerate}
  \item Initialize \(d=b=x=0\) and \(x_{\mathrm{old}}=\infty\), and calculate
  \[
    P=\left(\frac{1}{\alpha}A^{\mathsf T}A+\lambda I\right)^{-1}.
  \]
  \item During each outer iteration, repeat while
  \(\lVert x-x_{\mathrm{old}}\rVert_2\) exceeds the tolerance: set
  \(x_{\mathrm{old}}\leftarrow x\), and perform the prescribed inner updates.
  \item Update the reconstruction and auxiliary variables according to
  \[
    x\leftarrow P\!\left(A^{\mathsf T}y+\lambda(d-b)\right),\qquad
    d\leftarrow\operatorname{softshrink}\!\left(x+b,\frac{1}{\lambda}\right),
  \]
  followed by
  \[
    b\leftarrow b+\tau(x-d).
  \]
\end{enumerate}

\section{Results}

We first present CIFAR-10 reconstruction results, followed by a CIFAR-10-to-MNIST transfer test. We then examine sensitivity to Gaussian perturbations of the pattern matrix and display representative network-generated patterns. Finally, we discuss the scope and limitations of the results.

For training our neural network, we have used Amazon Web Services (AWS) using a specific EC2 instance with 48 virtual CPUs. The neural network was implemented using the PyTorch framework, which facilitated both the training and evaluation of the model.

\subsection{Reconstruction of similar images as a training set}

\begin{figure}[htbp]
    \centering
        \includegraphics[width=0.92\linewidth]{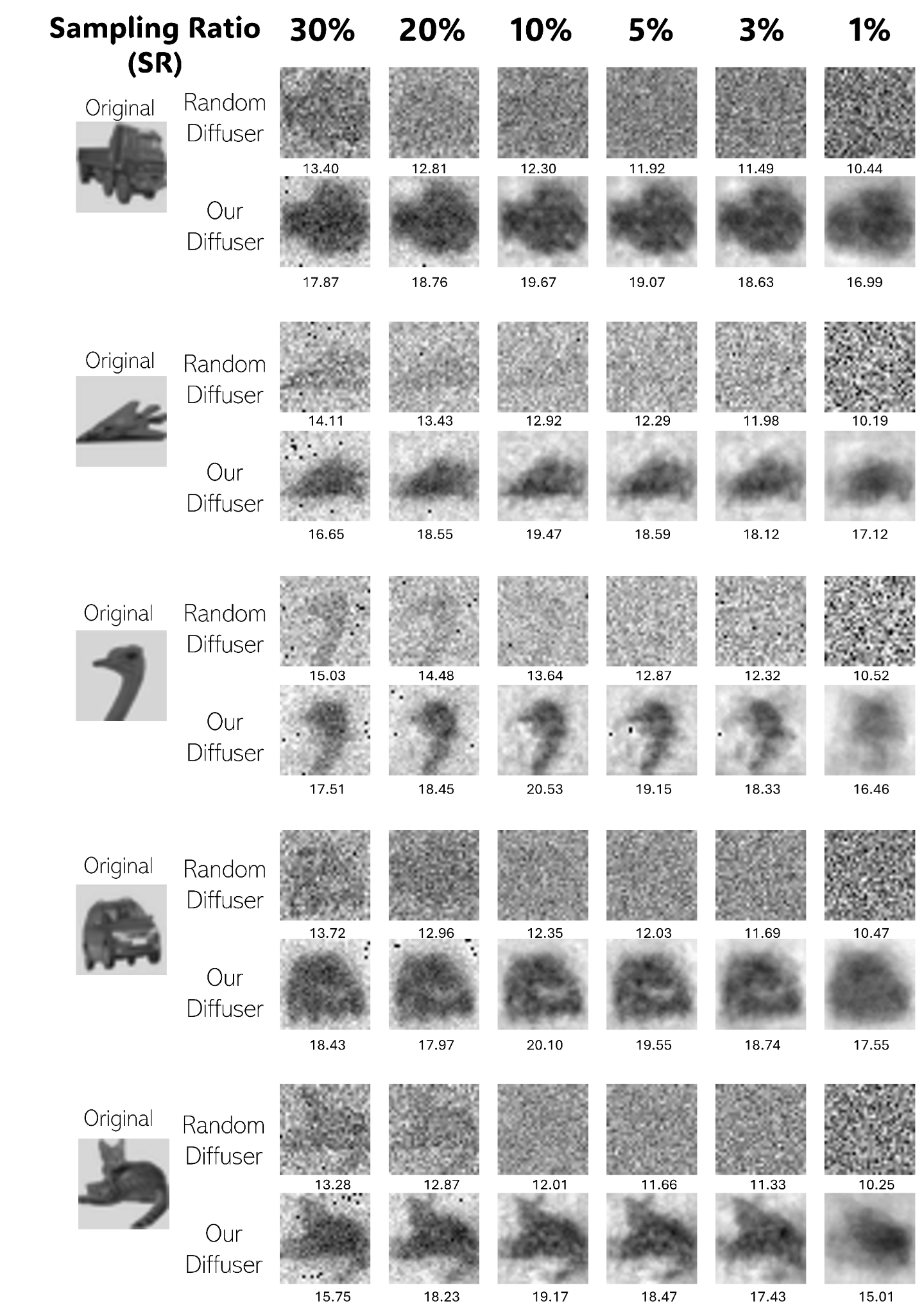}
    \caption{Image reconstruction obtained with both random and optimized diffusers for various objects and SRs. Comparison of image reconstructions using a random diffuser and our proposed diffuser across different SRs (30\%, 20\%, 10\%, 5\%, 3\%, and 1\%). The first column of each block shows the original image, followed by reconstructions using a random diffuser and using our method. The corresponding PSNR value is displayed below each reconstructed image, demonstrating the consistent advantage of our approach over the random diffuser approach, especially at lower SRs.}
    \label{fig: Image_reconstruction_comperesion}
\end{figure}

The comparison of reconstruction quality under varying sampling ratios (30\%, 20\%, 10\%, 5\%, 3\%, and 1\%) is illustrated in Fig. \ref{fig: Image_reconstruction_comperesion}. The distinction between the performance of a random diffuser and our proposed diffuser structure is evident across different levels of SRs. As the SR decreases, the reconstruction from a random diffuser exhibits more pronounced noise and reduced sharpness, with severe degradation at lower ratios, such as 3\% and 1\%. In contrast, our method consistently achieves more accurate reconstructions, maintaining discernible details even at lower sampling ratios.

Peak signal-to-noise ratio (PSNR) values displayed below each reconstructed image further highlight the advantage of our method over the random diffuser. Our approach yields higher PSNR values across all sampling ratios, with significant improvements at lower SRs like 1\% and 3\%, which are considered very small in the field. These results indicate that our network effectively adapts to reduced sampling ratios, providing enhanced image quality while the random diffuser struggles to preserve meaningful information under such conditions. The performance gap is especially notable as the sampling ratio drops below 10\%, where our method maintains better visual fidelity and structural details.
Beyond 10\%, the PSNR and SSIM values do not increase as expected, possibly because the network is not large enough to fully realize the potential of the method at higher sampling ratios. Nonetheless, the reconstruction performance with the network-based diffuser remains superior to that with the random diffuser.

We repeated the process with five different random diffuser sets to confirm that the observed improvements are not solely attributable to a particular suboptimal set of random diffusers. Similarly, in addition to training the network five times, for each set of final weights, we generated a different diffuser for each image to examine whether the same network weights lead to variations in reconstruction quality.

\begin{figure}[htbp]
    \centering
        \includegraphics[width=\linewidth]{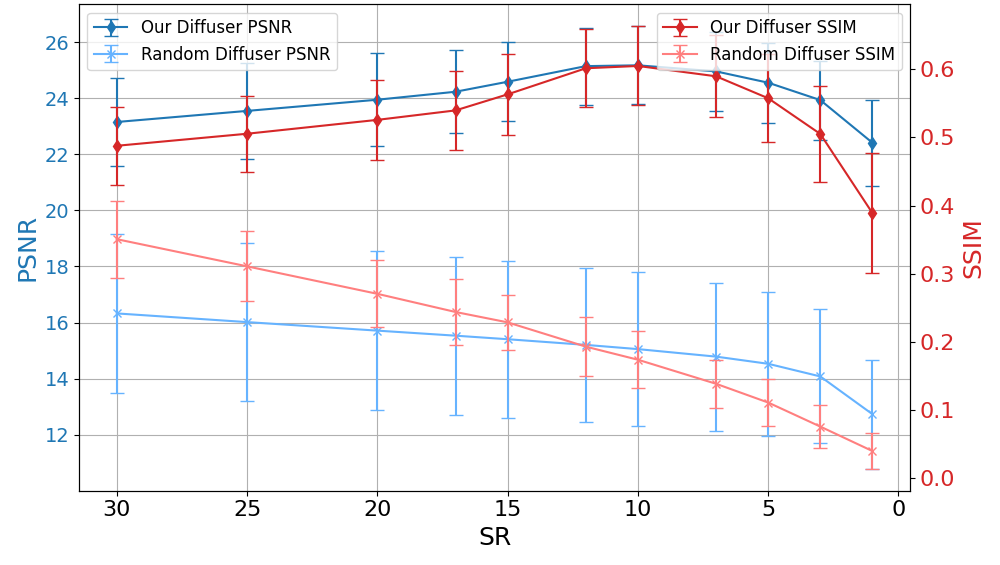}
    \caption{PSNR and SSIM as functions of SR. Comparison of PSNR and SSIM values for our neural network-based diffuser and 5 different sets of random diffusers across varying SRs. The left $y$-axis represents the PSNR value, while the right $y$-axis corresponds to the SSIM value. Our method consistently outperforms the random diffuser approach, maintaining higher PSNR and SSIM values, particularly at lower sampling ratios, indicating better image quality and structural similarity in the reconstructions.}
    \label{fig: PSNR_SSIM_vs_SR}
\end{figure}

The PSNR and structural similarity index measure (SSIM) results, presented in Fig. \ref{fig: PSNR_SSIM_vs_SR}, consistently show that our network outperforms the random diffuser, achieving significantly higher values across all SRs.

\subsection{Domain Adaptation}

Having obtained a useful set of illumination patterns after training on one dataset, we next test a limited form of cross-domain transfer. Generating large CGI training sets closely matched to every target object may be impractical, so it is useful to ask whether a pattern set learned from one image class retains value for another. This experiment is a CIFAR-10-to-MNIST transfer test, not evidence of universal generalization.

To explore this avenue, we tested our method using a test set containing different target objects than those in the training set. Specifically, we trained a network on the CIFAR-10 dataset and then evaluated how well the resulting diffuser set could reconstruct MNIST images that the network had not encountered during training.

\begin{figure}[htbp]
    \centering
        \includegraphics[width=\linewidth]{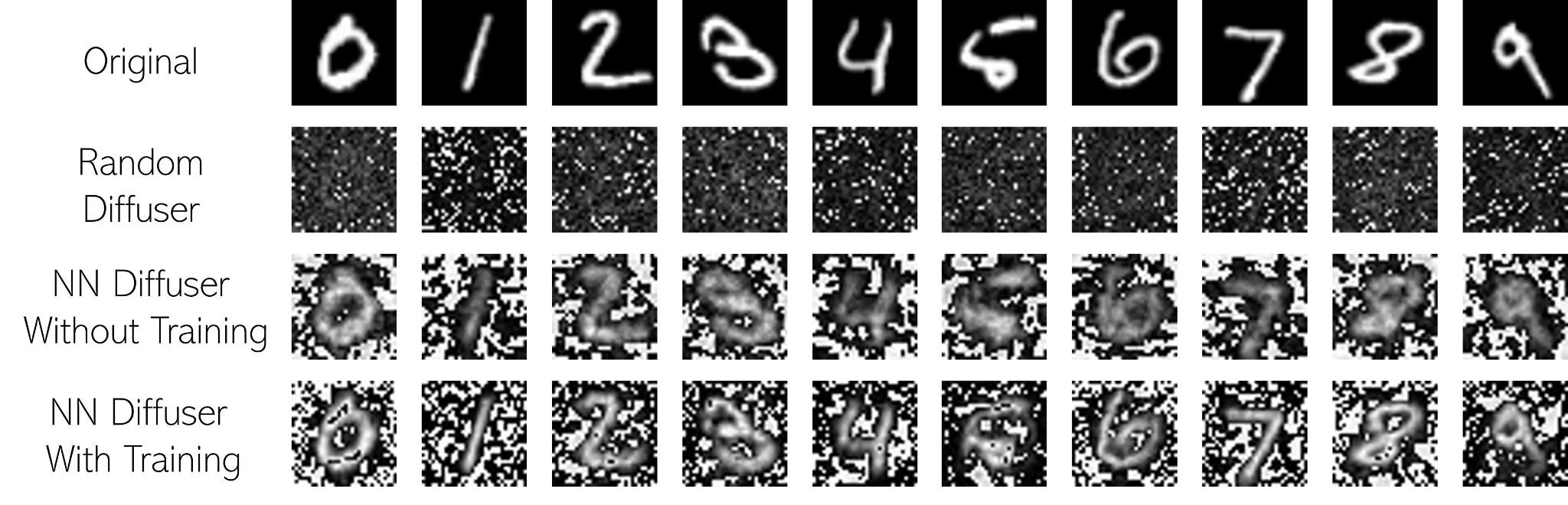}
    \caption{Image reconstruction using the previously generated diffusers for another dataset. Comparison of MNIST image reconstructions at SR = 10\%, using a random diffuser, a neural-network (NN) diffuser without training on MNIST, and an MNIST-trained NN diffuser. The top row shows the original digit images, followed by reconstructions using a random diffuser, which results in significant noise and distortion. The third row illustrates the results from the NN diffuser without MNIST training. The bottom row presents the output from the MNIST-trained NN diffuser.}
    \label{fig: Image_rec_mnist_sr_10}
\end{figure}

Figure \ref{fig: Image_rec_mnist_sr_10} illustrates the reconstruction quality of various methods applied to MNIST images in SR of 10\%. The top row displays the original images, followed by reconstructions using a random diffuser, a neural network-based diffuser without training on MNIST, and a neural network-based diffuser trained exclusively on MNIST images.

The random diffuser introduces significant noise, obscuring most details of the original digits. Meanwhile, the neural network-based diffuser, even without MNIST-specific training, produces recognizable and clear shapes. Although the best results are achieved with the diffuser trained on MNIST, the untrained version still delivers well-defined digit shapes and clearer details. This progression demonstrates that a training set does not necessarily need to include object-like images to achieve accurate image reconstructions.

\subsection{Sensitivity to Pattern Perturbations}

We assess sensitivity to Gaussian perturbations of the learned pattern matrix using 18 CIFAR-10 images. Specifically, the perturbation is applied elementwise to \(\boldsymbol{A}\), corresponding to \(\Delta\boldsymbol{A}\) in Eq. \ref{eq:measurement_model}. This models pattern-generation or calibration error. It does not model additive detector noise, read noise, or photon-counting noise, which would enter through \(\boldsymbol{n}\).

\begin{figure}[htbp]
    \centering
        \includegraphics[width=0.82\linewidth]{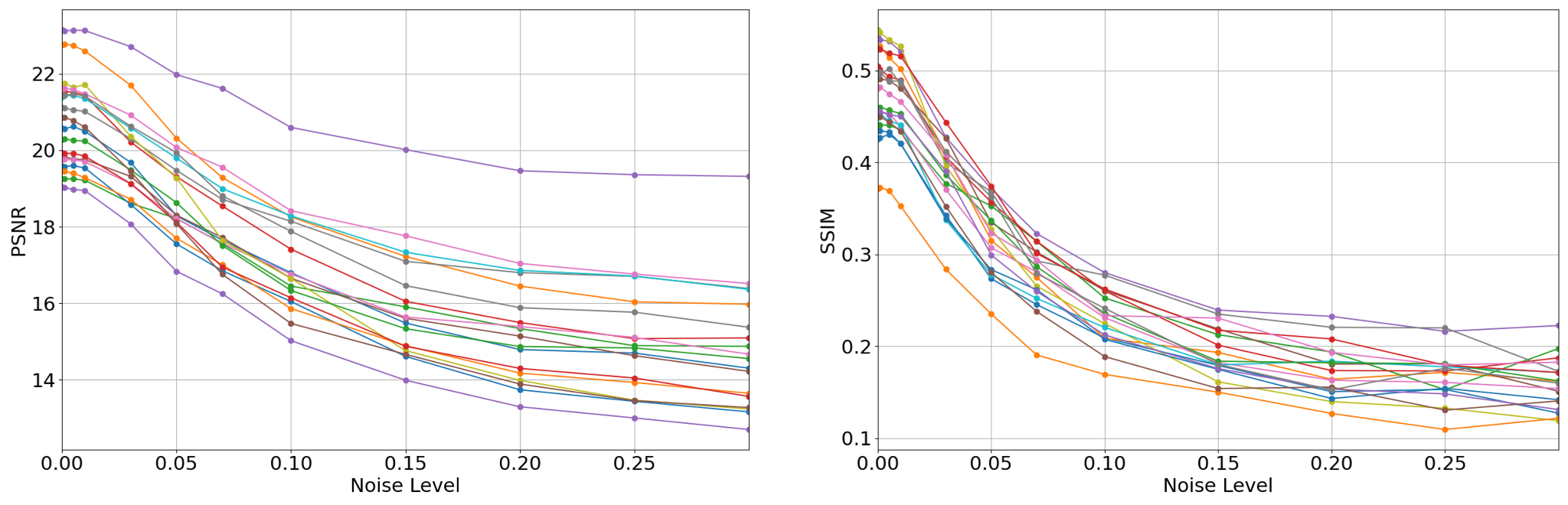}
    \caption{Sensitivity to pattern-matrix perturbations. PSNR and SSIM of the reconstructed images as functions of the elementwise Gaussian perturbation applied to the diffuser matrix. This test represents pattern or calibration error, not additive detector or photon-counting noise.}
    \label{fig: noise_exp}
\end{figure}

At a fixed SR of 10\%, Fig. \ref{fig: noise_exp} shows little change in PSNR or SSIM for a 1\% matrix perturbation, whereas performance degrades substantially as the perturbation approaches 30\%. The result therefore supports local tolerance to small pattern errors under the simulated conditions, rather than a general claim of noise robustness.

\subsection{Diffuser Sets}

\begin{figure}[htbp]
\centering
        \includegraphics[width=\linewidth]{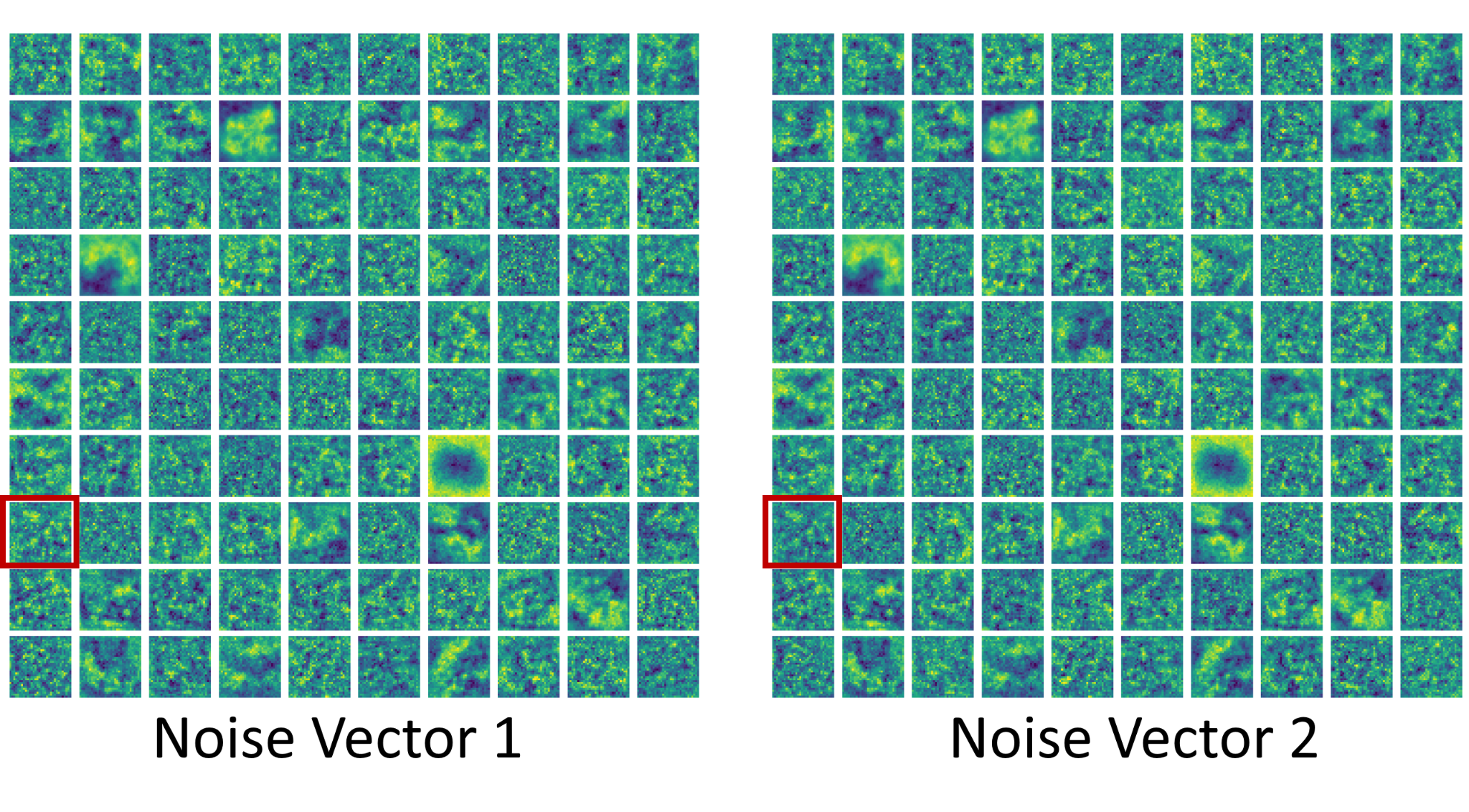}
    \caption{Comparison of different sets of diffuser patterns in SR=10\% corresponding to two different noise vectors. The patterns are very similar, but not identical, having features which reflect both unique structure and randomness.}
    \label{fig: patterns}
\end{figure}

Figure \ref{fig: patterns} presents 100 diffuser patterns generated after optimization, each with \(32\times32\) pixels and together corresponding to an SR of about 10\%. Under the tested reconstruction procedure, these patterns yield better image-quality metrics than the random-pattern baseline. Different input noise vectors supplied to the trained generator produce pattern sets with notable similarities.

Although the patterns may appear identical at first glance, closer inspection reveals intensity differences within a recurring overall structure. Some features visually resemble familiar signal-processing filters, but establishing such a connection requires quantitative analysis of spectra, correlations, and conditioning. This provides a useful direction for future work in compressed sensing and structured illumination.

\subsection{Discussion}


The learned diffuser sets display a nontrivial recurring structure and improve PSNR and SSIM relative to the tested random baseline. This advantage appears at every evaluated SR and is especially visible at 3\% and 1\%. The result establishes that jointly adapting the pattern ensemble to the training distribution and reconstruction loss can benefit strongly undersampled CGI. It does not yet establish an advantage over established structured or optimized pattern families.

Repeating the comparison with five random diffuser sets and five training runs reduces the likelihood that the observed gain is caused by a single unfavorable random draw. The perturbation study further indicates tolerance to small errors in the realized pattern matrix, followed by substantial degradation for large errors. Because additive detector and photon-counting noise were not simulated, these results should not be interpreted as full system-level robustness.

Patterns trained on CIFAR-10 also produce recognizable MNIST reconstructions and outperform the random baseline in this transfer test. This is encouraging evidence that the learned patterns are not restricted to reproducing the training images. Since CIFAR-10 and MNIST constitute only one source--target pair, broader transfer claims require additional object classes and physical measurements.

\section{Conclusion}

We presented a deep-learning approach for designing grayscale illumination patterns for computational ghost imaging at low sampling ratios. Rather than replacing the reconstruction stage with a neural network, the method learns the sensing patterns and then uses Split Bregman reconstruction.

In the reported simulations, the learned patterns improve PSNR and SSIM relative to random patterns across the tested SRs, with recognizable reconstructions obtained down to 1\%. A pattern set trained on CIFAR-10 also transfers usefully to MNIST, and small perturbations of the sensing matrix cause little degradation under the tested conditions.

These findings support learned pattern design as a promising way to reduce the number of CGI measurements while retaining reconstruction quality. They do not establish global optimality, superiority to structured patterns, robustness to detector or photon noise, or a reduction in total radiation dose. Those questions require matched-fluence comparisons and experimental validation.

The most direct next steps are comparisons with Hadamard, Fourier/DCT, wavelet, and other learned or optimized pattern ensembles; evaluation under additive Gaussian and Poisson bucket noise at equal photon budgets; binary or quantized mask constraints; and a physical CGI demonstration. The architecture can also be retrained around other reconstruction algorithms, but that generality should be tested rather than assumed.

\begin{backmatter}
\bmsection{Funding}
European Union's Horizon Europe research and innovation programme (101178170); Israel Science Foundation (2208/24).

\bmsection{Acknowledgment}
We wish to thank Ethan Fetaya, Or Sefi, Sharon Shwartz, and Amotz Taub-Tabib for helpful comments and discussions.

\bmsection{Disclosures}
The authors declare no conflicts of interest.

\bmsection{Data availability}
Data underlying the results presented in this paper are not publicly available at this time but may be obtained from the authors upon reasonable request.
\end{backmatter}

\end{document}